\documentstyle[epsfig,rotate]{ioplppt}

\newcommand{\gccm}{\rm\,g\,cm^{-3}}

\begin{document}

International Symposium on ''Strangeness in Quark Matter 1997``, April
14--18, Thera (Santorini), Hellas

\jl{4}
\title{Are strange stars distinguishable from neutron stars by their 
       cooling behaviour?}
\author{Ch Schaab\ftnote{1}
        {E-mail address: schaab@gsm.sue.physik.uni-muenchen.de},
        B Hermann, F Weber and M K Weigel}
\address{Institut f{\"u}r theoretische Physik,
  Ludwig-Maximilians Universit{\"a}t M{\"u}nchen, Theresienstr. 37,
  D-80333 M{\"u}nchen, Germany}

\begin{abstract}
  The general statement that strange stars cool more rapidly than
  neutron stars is investigated in greater detail. It is found that
  the direct Urca process could be forbidden not only in neutron stars
  but also in strange stars. If so, strange stars would be slowly
  cooling and their surface temperatures would be more or less
  indistinguishable from those of slowly cooling neutron stars.  The
  case of enhanced cooling is reinvestigated as well.  It is found
  that strange stars cool significantly more rapidly than neutron
  stars within the first $\sim 30$ years after birth. This feature
  could become particularly interesting if continued observation of SN
  1987A would reveal the temperature of the possibly existing pulsar
  at its centre.
\end{abstract}


\maketitle

%%%%%% 1. Introduction 
\section{Introduction}
The theoretical possibility that strange quark matter -- made up of
roughly equal numbers of up, down and strange quarks -- may be more
stable than atomic nuclei (specifically iron, which is the most stable
atomic nucleus) constitutes one of the most startling predictions of
modern physics
\cite{Bodmer71,Witten84,Terazawa89b,Terazawa89a,Terazawa90a}, which,
if true, would have implications of greatest importance for laboratory
physics, cosmology, the early universe, its evolution to the present
day, and massive astrophysical objects \cite{Aarhus91}.  Unfortunately
it seems unlikely that lattice QCD calculations will be accurate
enough in the foreseeable future to give a definitive prediction on
the absolute stability of strange matter, so that one is presently
left with experiments and astrophysical studies
\cite{Glendenning92,Glendenning94a,Glendenning94b} to either confirm
or reject the absolute stability of strange matter. In a recent
investigation \cite{Schaab97b}, dealing with the second item, we
compared the cooling behaviour of neutron stars with the one of their
hypothetical strange counterparts -- strange stars
\cite{Witten84,Haensel86,Alcock86,Glendenning90}.  The theoretical
predictions were compared with the body of observed data taken by
ROSAT and ASCA. There have been investigations on this topic prior to
this one (e.g., see \cite{Alcock88,Pizzochero91,Page91a,Schaab95a}).
These, however, did not incorporate the so-called standard cooling
scenario that turns out to be possible not only in neutron star matter
but in strange quark matter too, altering some of the conclusions made
in the earlier investigations significantly.

In the following section we will shortly review the structure of
strange stars in comparison to neutron stars. Since the composition of
strange matter turns out to be of great importance for the neutrino
emissivity, we shall consider its determination in greater detail in
section \ref{sec:eos}. The various neutrino emission processes and
observational data are discussed in sections \ref{sec:emissivity} and
\ref{sec:observations}, respectively. In the last section, we present
and discuss the results of cooling simulations and compare them with
observed data.

%%%%%% Structure
\section{Structure of strange stars and neutron stars}

The cross section of a neutron star can be devided roughly into four
distinct regimes. The outermost layer, with an optical depth of $\tau
\sim 1$, is called photosphere. The thermal radiation, which can be
observed by X-ray telescopes, is emitted from this region. This
radiation dominates the cooling of pulsars older than $\sim
10^6$~yrs. The second regime is the star's outer crust, which consists
of a lattice of atomic nuclei and a Fermi liquid of relativistic,
degenerate electrons.  The outer crust envelopes what is called the
inner crust, which extends from neutron drip density, $\rho=4.3\times
10^{11}\gccm$, to a transition density of about $\rho_{\rm
tr}=1.7\times 10^{14}\gccm$ \cite{Pethick95}. Neutrons, both inside
and outside of the atomic nuclei, are believed to form superfluid
cooper pairs in this regime.

Beyond $\rho_{\rm tr}$ one enters the star's fourth regime, that is,
its core where all atomic nuclei have dissolved into their
constituents, protons, neutrons, and -- due to the high Fermi pressure
-- possibly hyperons, more massive baryon resonances, and up, down and
strange quarks. The latter possibility leads to so-called hybrid stars
\cite{Glendenning97a}, which are to be distinguished from strange
stars made up of absolutely stable atrange quark matter.  Finally
meson condensates may be found in the core, too.  Neutrons and protons
may form superfluid states in the core of a neutron star. It is
however questionable whether or not the superfluids reach to the
centre of the star. Solutions of the gap equation differ in the gap
energy by almost one order of magnitude (see, for instance,
\cite{Amundsen85,Elgaroy96c}).  To account for these uncertainties, we
shall investigate models with and without superfluid cores.

In the first $\sim 10^6$~yrs the cooling of a neutron star is dominated
by emission of neutrinos from the core. Depending on the possible
neutrino reactions one can distinguish between standard and enhanced
cooling (see section \ref{sec:emissivity}). Both the inner and 
outer crust act as a thermal insulator between the cooling core and
the surface.

Since absolutely stable strange quark matter is selfbound, gravity is
not necessary to bind strange stars, in contrast to neutron stars. As
pointed out by Alcock, Farhi, and Olinto \cite{Alcock86}, a strange
star can carry a solid nuclear crust whose density at its base is
strictly limited by neutron drip.  This is made possible by the
displacement of electrons at the surface of strange matter, which
leads to a strong electric dipole layer there. It is sufficiently
strong to stabilize a gap between ordinary atomic (crust) matter and
the quark-matter surface, which prevents a conversion of ordinary
atomic matter into the assumed lower-lying ground state of strange
matter.  Obviously, free neutrons, being electrically charge neutral,
cannot exist in the crust, because they do not feel the Coulomb
barrier and thus would gravitate toward the strange-quark matter core,
where they are converted by hypothesis into strange matter.
Consequently, the density at the base of the crust (inner crust
density) will always be smaller than neutron drip density.  The main
differences with respect to the structure of a neutron star is the
composition of the core and the absence of the inner crust in the case
of strange stars. As we will see the latter results in a smaller
thermal insulation between the core and the surface of a strange star
than in the case of a neutron star
\cite{Alcock88,Pizzochero91}.

%%%%%% Composition
\section{Description of strange matter} \label{sec:eos}

We use the MIT bag model including $O(\alpha_{\rm{s}})$-corrections
\cite{Chodos74,Farhi84} to model the properties of absolutely stable
strange matter. Its equation of state and
quark-lepton composition, which is governed by the conditions of
chemical equilibrium and electric charge neutrality, is derived for
that range of model parameters -- that is, bag constant $B^{1/4}$, the
strange quark mass $m_{\rm{s}}$, and strong coupling constant
$\alpha_{\rm{s}}$ -- for which strange matter is absolutely stable
(i.e. energy per baryon $E/A$ less than the one of $^{56}$Fe,
$E/A=930$~MeV)  (see also \cite{Schertler97b} where a different result
for the composition was obtained).

\begin{figure} \begin{indented} \item[]
\psfig{figure=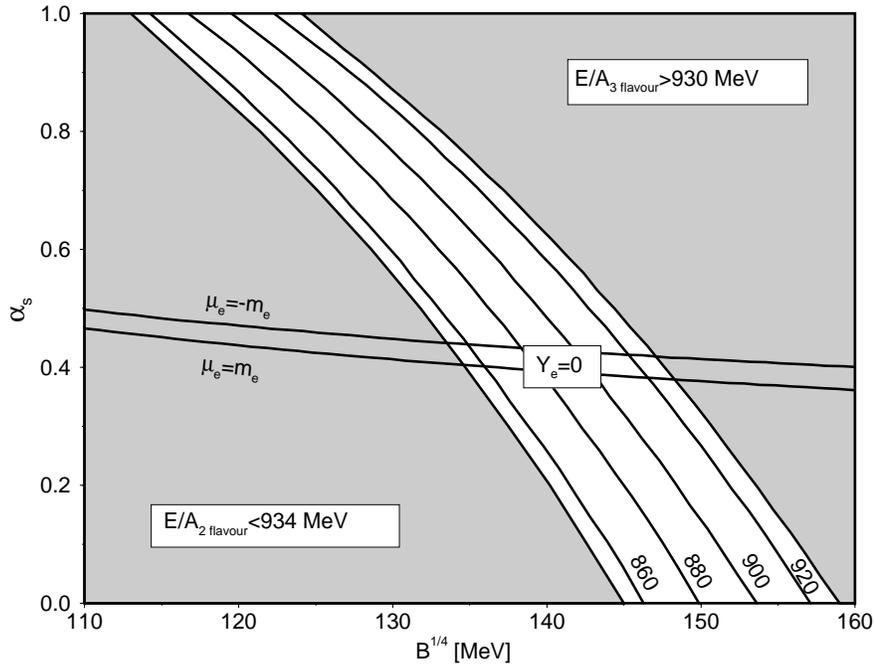,height=0.8\linewidth,angle=-90}
\end{indented}
\caption[]{
  Allowed parameter range for which strange quark matter is absolutely
  stable.  The shaded regions represent regions where either three
  flavour quark matter is not absolutely stable or nucleons would
  decay into quark matter. The labels attached to the four diagonal
  curves give the energy per baryon (in MeV) of three flavour quark
  matter. The two almost horizontal lines seperate the region where no
  electrons exist.
  \label{fig:parameter}}
\end{figure}

In the limiting case of vanishing quark masses, electrons are not
necessary to achieve charge neutrality. In the more realistic case of a
finite strange quark mass $m_{\rm s}$, the electrons can nevertheless
vanish above a certain density, which depends on $\alpha_{\rm s}$. Figure
\ref{fig:parameter} shows the allowed parameter space of $\alpha_s$
and $B^{1/4}$ for a fixed strange quark mass of $m_{\rm s}=100$~MeV. This
space is limited by two constraints. Firstly, the energy per baryon of
three flavour quark matter has to be less than the one of iron
(930~MeV), secondly, the energy per baryon of two flavour (up- and
down-) quark matter has to be above the one of nucleons (938~MeV)
minus a surface energy correction ($\sim 4$~MeV) \cite{Farhi84}. The
almost horizontal lines represent the parameter sets for which the
chemical potential of electrons and positrons are equal to
their rest masses. In the region between these two lines electrons and
positrons disappear even in the case of a nonvanishing strange quark
mass.  The behaviour of the electron's chemical potential depends on
the chosen renormalization point $\rho$. We followed Duncan et al.
\cite{Duncan83} by renormalizing on shell ($\rho =m_{\rm s}$). The
renormalization $\rho=300~{\rm MeV}~(\approx\mu_{\rm s})$ suggested by
Farhi and Jaffe \cite{Farhi84} reduces the strangeness fraction and
thus enhances the electron's chemical potential. The $Y_{\rm
e}=0$-region is therefore shifted to higher $\alpha_{\rm
s}$-values. Since the MIT bag model is only phenomenlogical, it seems
presumptous to draw definitive conclusions for both cases.

It was pointed out by Duncan et al. \cite{Duncan83} (see also
\cite{Alcock86,Pethick92a}) that the neutrino emissivity of strange
matter depends strongly on its electron fraction, $Y_{\rm e}$. For
that reason we introduce two different, complementary parameter sets
denoted SM-1 and SM-2, which correspond to strange matter that contains
a relatively high electron fraction (SM-1, below the bottom line in
figure \ref{fig:parameter}), and $Y_{\rm e}=0$ (SM-2, between the two
lines) for the density range of interest here.

%%%%%% 3. NEUTRINO EMISSIVITY

\section{Neutrino emissivity} \label{sec:emissivity}
The neutrino emission processes can be divided into slow and fast ones
(see table \ref{tab:neutrino} for the most important reactions in
the cores of neutron and strange stars). The large
difference in the emissivities is caused by the rather different phase
spaces associated with these reactions. The available phase space of
the slow reactions is that of a two-baryon scattering process, whereas
it is that of a one-baryon decay process for the fast reactions.  The
only fast processes in quark matter (the quark {\em direct} Urca
processes)
\begin{equation}
  \rm{d} \rightarrow \rm{u} + \rm{e}^- + \bar\nu_{\rm{e}} \label{eq:dtou}
\end{equation}
and
\begin{equation}
  \rm{s} \rightarrow \rm{u} + \rm{e}^- + \bar\nu_{\rm{e}}, \label{eq:stou}
\end{equation}
as well as their inverse reactions are only possible if the fermi momenta
of quarks and electrons ($p_{\rm{F}}^i$, $i={\rm u,d,s;e}^-$) fulfill
the so-called triangle inequality (e.g.,
$p_{\rm{F}}^{\rm{d}}<p_{\rm{F}}^{\rm{u}}+p_{\rm{F}}^{\rm{e}}$ for
process (\ref{eq:dtou})). This relation is the analogue to the
triangle inequality established for nucleons and electrons in the
nuclear matter case (nucleon direct Urca process
\cite{Boguta81a,Lattimer91}).

\begin{table}
\caption{The most important neutrino emission processes in the cores
  of neutron stars and strange stars. The associated emissivities and
  constraints for the respective process are also given.
  \label{tab:neutrino}}
\footnotesize \rm \begin{tabular}{@{}lccc}
\br
\centre{1}{Process} 	& Emissivity	& Rapidity	& Constraints \\
\mr
\centre{4}{Neutron stars} \\ 
\mr
modified Urca, e.g. 
${\rm n}+{\rm p}+{\rm e}^- \rightarrow {\rm n}+{\rm n}+\nu_{\rm e}$
& $\sim 10^{21}\times T_9^8$	& slow	& \\
direct Urca, e.g.
${\rm p}+{\rm e}^- \rightarrow {\rm n}+\nu_{\rm e}$
& $\sim 10^{28}\times T_9^6$	& fast	& $Y_{\rm p} > 0.11$ \\
K-, $\pi$-condensation
& $\sim 10^{24-25}\times T_9^6$	& fast	& $n>n_{\rm c}\sim 3-5 n_0$ \\
\mr
\centre{4}{Strange stars} \\
\mr
modified Urca, e.g. 
${\rm d}+{\rm u}+{\rm e}^- \rightarrow {\rm d}+{\rm d}+\nu_{\rm e}$
& $\sim 10^{20}\times T_9^8$	& slow	& $Y_{\rm e}>0$ \\
direct Urca, e.g. 
${\rm u}+{\rm e}^- \rightarrow {\rm d}+\nu_{\rm e}$
& $\sim 10^{24}\times T_9^6$	& fast	& $Y_{\rm e} > 0$ \\
Bremsstrahlung
$q_1+q_2 \rightarrow q_1+q_2+\nu+\bar\nu$
& $\sim 10^{19}\times T_9^8$	& slow	&  \\
\br
\end{tabular} \end{table}

If the electron fermi momentum is too small (i.e., $Y_{\rm e}$ is too
little), then the triangle inequality for the processes
(\ref{eq:dtou}) and (\ref{eq:stou}) cannot be fulfilled and a
bystander quark is needed to ensure energy and momentum conservation
in the scattering process. The latter process is known as the quark
{\em modified} Urca process, whose emissivity is considerably smaller
than the emissivity of the direct Urca process.  If the electron
fraction vanishes entirely, as is the case for SM-2, both the quark
direct and the quark modified Urca processes become unimportant. The
neutrino emission is then dominated by bremsstrahlung processes only,
\begin{equation}
  Q_1 + Q_2 \longrightarrow Q_1 + Q_2 + \nu + \bar\nu,
\end{equation} 
where $Q_1$, $Q_2$ denote any pair of quark flavours.  For the
emissivities associated with the quark direct Urca, quark modified
Urca, and quark bremsstrahlung processes, we refer to 
references \cite{Duncan83,Price80a,Iwamoto82}.

It has been suggested \cite{Bailin79a,Bailin84} that the quarks
eventually may form Cooper pairs. This would suppress, as in the
nuclear matter case, the neutrino emissivities by an exponential
factor of $\exp(-\Delta/k_{\rm{B}}T)$, where $\Delta$ is the gap
energy, $k_{\rm{B}}$ Boltzmann's constant, and $T$ the temperature.
Unfortunately, up to now there exists neither a precise experimentally
nor theoretically determined value for the gap energy.  In order to
give a feeling for the influence of a possibly superfluid behaviour of
the quarks in strange matter, we choose $\Delta=0.4$~MeV, as estimated
in the work of Bailin and Love \cite{Bailin79a}. (Such a $\Delta$
value is not too different from the nuclear-matter case, where the
proton $^1{\rm S}_0$ gap, for instance, amounts $\sim 0.2 {\rm -}
1.0$~MeV \cite{Elgaroy96c,Wambach91a}, depending on the
nucleon-nucleon interaction and the microscopic many-body model.) The
outcome of our superfluid strange matter calculations will be labeled
SM-1$^{\rm{sf}}$ and SM-2$^{\rm{sf}}$.

%%%%%%% OBSERVED DATA
\section{Observed data} \label{sec:observations}
Among the soft X-ray observations of the 23 sources which were
identified as pulsars, the ROSAT and ASCA observations of PSRs
0002+62, 0833-45 (Vela), 0656+14, 0630+18 (Geminga) and 1055-52 (see
table \ref{tab:observations}) achieved a sufficiently high photon flux
such that the effective surface temperatures of these pulsars could be
extracted by two- or three-component spectral fits \cite{Oegelman95a}.
The obtained effective surface temperatures, shown in figures
\ref{fig:nsf} and \ref{fig:sf}, depend crucially on whether a hydrogen
atmosphere is used or not.  Since the photon flux measured solely in
the X-ray energy band does not allow one to determine what kind of
atmosphere one should use, we consider both the blackbody model and
the hydrogen-atmosphere model, drawn in in Figs.  \ref{fig:nsf} and
\ref{fig:sf} as error bars with a solid and hollow circle.  The kind
of atmosphere of individual pulsars could be determined by considering
multiwavelength observations \cite{Pavlov96a}. All error bars
represent the $3\sigma$ error range due to the small photon fluxes.

\begin{table}
\caption[]{Surface temperatures as measured by an observer at 
  infinity, $T_{\rm s}^\infty$, and spin-down ages, $\tau$, of
  observed pulsars. \label{tab:observations}}
\begin{indented} \item[] \begin{tabular}{@{}ccccc}
\br
Pulsar & $\log\tau$ [yrs] & Model atmosphere & $\log T_{\rm s}^\infty$ [K] 
& Reference \\
\mr
0002+62 & $\sim 4$\dag 		& blackbody             
& $6.20^{+0.09}_{-0.40}$        & \cite{Hailey95a} \\
\mr
0833-45 & $4.4\pm 0.1$\dag	& blackbody             
& $6.24\pm 0.08$                & \cite{Oegelman95a} \\
(Vela)  &                  	& magnetic H-atmosphere 
& $5.88\pm 0.06$ 		& \cite{Page96a} \\
\mr
0656+14 & $5.05$ 		& blackbody
& $5.89^{+0.08}_{-0.33}$ 	& \cite{Greiveldinger96a} \\
        &                  	& magnetic H-atmosphere 
& $5.72^{+0.06}_{-0.03}$	& \cite{Anderson93} \\
\mr
0630+18 & $5.53$ 		& blackbody
& $5.75^{+0.05}_{-0.08}$        & \cite{Halpern97a} \\
(Geminga)&                 	& H-atmosphere 
& $5.42^{+0.12}_{-0.04}$        	& \cite{Meyer94} \\
\mr
1055-52 & $5.73$ 		& blackbody
& $5.90^{+0.09}_{-0.21}$ 	& \cite{Greiveldinger96a} \\
\br
\end{tabular} 
\item[]
\dag~~ estimated true age instead of spin-down age (see text).
\end{indented}
\end{table}

Except for PSRs 0833-45 (Vela) and 0002+62, all ages are
estimated by their spin-down age $\tau=P/2\dot P$. This relation
implies however that both the moment of inertia and the magnetic surface
field are constant with time, and that the braking index $n$ is equal to its
canonical value 3 (angular momentum loss dure to pure magnetic dipole
radiation). The true ages may therefore be quite different from the
spin-down ages. The age of Vela was recently determined
in reference \cite{Lyne96a}, and the approximate age of PSR 0002+62 is
given by an estimate of the age of the related supernova remnant G
117.7+06.

%%%%%% NUMERICAL SIMULATION

\section{Results and Discussion} 
\begin{figure} \begin{indented} \item[]
\psfig{figure=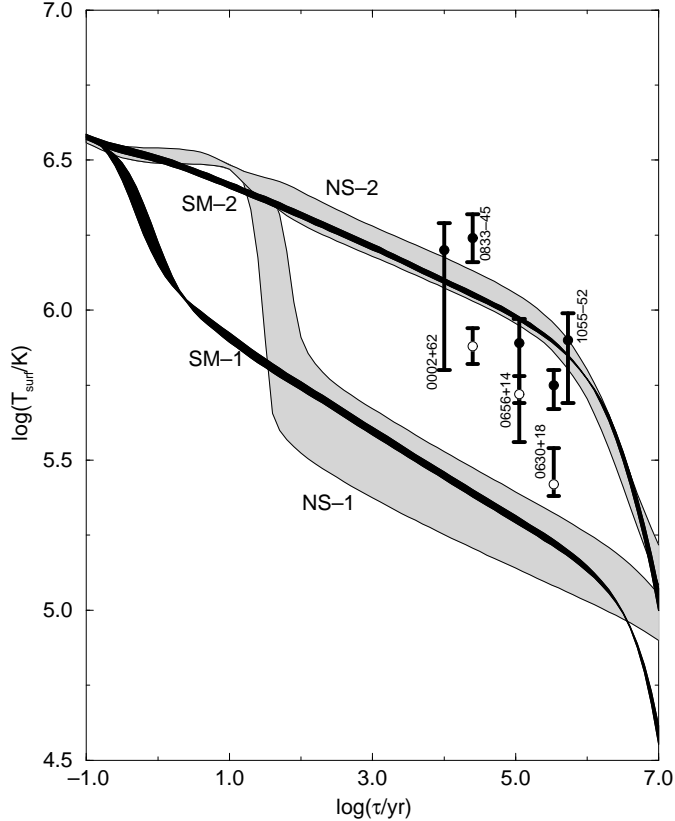,width=0.8\linewidth}
\end{indented}
\caption[]{
  Cooling of non-superfluid strange star models SM-1 (lower solid
  band) and SM-2 (upper solid band), and neutron star models NS-1
  (lower dotted band) and NS-2 (upper dotted band). The surface
  temperatures obtained with a blackbody- (magnetic hydrogen-)
  atmosphere are marked with error bars with solid (hollow) circle
  representing the most probable values (see table
  \ref{tab:observations}). \label{fig:nsf}}
\end{figure}

\begin{figure}\begin{indented} \item[] 
\psfig{figure=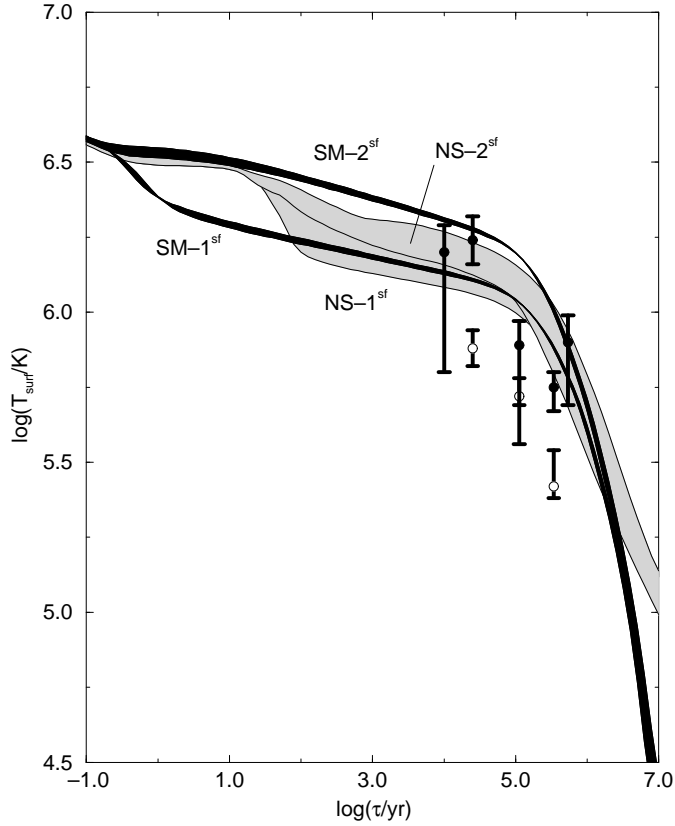,width=0.8\linewidth}
\end{indented}
\caption[]{
  Cooling of superfluid strange star SM-1$^{\rm sf}$ (lower solid 
  band) and SM-2$^{\rm sf}$ (upper solid band), and neutron star
  models NS-1$^{\rm sf}$ (lower dotted band) and NS-2$^{\rm sf}$ (upper
  dotted band). \label{fig:sf}}
\end{figure}

The thermal evolution of strange stars and neutron stars was simulated
using the evolutionary numerical code described in Schaab et al.
\cite{Schaab95a} (see also
\cite{Tsuruta66,Richardson82,VanRiper91,Page95,Schaab95b}). The
neutron star models are based on a broad collection of EOSs which
comprises relativistic, fieldtheoretical equations of state as well as
non-relativistic, Schroedinger-based ones (see \cite{Schaab95a} for
details).  As a specific feature of the relativistic models, they
account for all baryon states that become populated in dense neutron
star matter up to the highest densities reached in the cores of the
heaviest neutron stars constructed from this collection of equations
of state.  Neutron stars are known to loose energy either via standard
cooling or enhanced cooling.  Both may be delayed by a superfluid
core.  Consequently all four options are taken into account
here. These are labeled in figures \ref{fig:nsf} and \ref{fig:sf} as
NS-1 (enhanced cooling) and NS-2 (standard cooling) for normal neutron
star matter, and NS-1$^{\rm{sf}}$ and NS-2$^{\rm{sf}}$ (delayed
cooling) for superfluid neutron star matter.  The parameters of
NS-1$^{\rm{sf}}$ and NS-2$^{\rm{sf}}$ are listed in table~4 of
reference \cite{Schaab95a}.  In analogy to this, the corresponding
strange-star cooling curves are SM-1 (enhanced cooling) and SM-2
(standard cooling) for normal strange quark matter, and
SM-1$^{\rm{sf}}$ and SM-2$^{\rm{sf}}$ (delayed cooling) for superfluid
quark matter.


All calculations are performed for a gravitational star mass of
$M=1.4M_\odot$, about which the observed pulsar masses tend to
scatter. The band-like structure of the cooling curves is supposed to
reflect the uncertainties inherent in the equation of state of
neutron-star and strange-star matter.  These have their origin, in the
case of neutron stars (gray bands), in the different many-body
techniques used to solve the nuclear many-body problem.  In the latter
case, strange-star matter, the solid bands refer to the range of
allowed bag values, $B^{1/4}$ (see figure
\ref{fig:parameter}), for which strange quark matter is absolutely
stable.  One might suspect that the large gap between the cooling
tracks of SM-1 and SM-2 in figure \ref{fig:nsf} can be bridged
steadily by varying the strong coupling constant $\alpha_{\rm s}$.
However it turns out that this gap can be filled only for $\alpha_{\rm
  s}$-values that lie within an extremely small range. This is caused
by the sensitive functional relationship between $\alpha_{\rm s}$ and
the neutrino luminosity $L_\nu$, which is rather steep around that
$\alpha_{\rm s}$-value for which the electrons vanish from the quark
core of the star. All other values of $\alpha_{\rm s}$ give cooling
tracks which are close to the upper or lower bands. This behaviour
resembles the case of neutron stars, where the neutrino luminosity
depends sensitively on the star's mass.

One sees from Figs. \ref{fig:nsf} and \ref{fig:sf} that, except for
the first $\sim 30$ years of the lifetime of a newly born pulsar, both
neutron stars and strange stars may show more or less the {\em same}
cooling behaviour, provided both types of stars are made up of either
normal matter or superfluid matter.  (We will come back to this issue
below.) This is made possible by the fact that both standard cooling
(NS-2) as well as enhanced cooling (NS-1) in neutron stars has its
counterpart in strange stars too (SM-2 and SM-1, respectively).  The
point of time at which the surface temperature drop of a strange star
occurs depends on the thickness of the nuclear crust that may envelope
the strange matter core and thermaly insulate it from the surface
\cite{Schaab95a}. In the present calculation, strange stars possess
the densest possible nuclear crust, which is about 0.2 km thick.
Thinner crusts would lead to temperature drops at even earlier times.
Figures \ref{fig:nsf} and \ref{fig:sf} indicate that the cooling data
of observed pulsars do not allow to decide about the true nature of
the underlying collapsed star, that is, as to whether it is a strange
star or a conventional neutron star.  This could abruptly change with
the observation of a very young pulsar shortly after its formation in
a supernova explosion.  In this case a prompt drop of the pulsar's
temperature, say within the first 30 years after its formation, could
offer a good signature of a strange star \cite{Alcock88,Pizzochero91}.
This feature, provided it withstands a rigorous future analysis of the
microscopic properties of quark matter, could become particularly
interesting if continued observation of SN 1987A would reveal the
temperature of the possibly existing pulsar at its centre.

Finally, we add some comments about the possibility that only the
neutron star is made up of superfluid matter but not the strange star.
In this case one has to compare the models SM-1 and SM-2 (see
figure \ref{fig:nsf}) with models NS-1$^{\rm sf}$ and NS-2$^{\rm sf}$
(see figure \ref{fig:sf}) yielding to an overall different cooling
history of neutron stars and enhanced-cooling strange stars
(SM-1). Therefore, the standard argument pointed out quite frequently
in the literature that strange stars cool much more rapidly than
neutron stars applies only to this special case.

%%%% Acknowledgments

\ack
We would like to thank J. Madsen for his comments.
  
%%%%%%%%%%%%%%%%%%%%%%%%%%%%%%%% REFERENCES %%%%%%%%%%%%%%%%%%%%%%%%%%%%
\section*{References}

%\bibliography{../bibtex/diplom}
%\bibliographystyle{../bibtex/nucphys}

\end{document}